\documentstyle[11pt,nature]{article}
\lefthead{de Bernardis et al.}
\righthead{}

\begin{document}
\author{\raggedleft In April 27th 2000 issue of Nature}

\title{A Flat Universe from High-Resolution Maps of the Cosmic 
Microwave Background Radiation\\}

\author{
\raggedright P. de Bernardis\altaffilmark{1},
P.A.R.Ade\altaffilmark{2},  
J.J.Bock\altaffilmark{3}, 
J.R.Bond\altaffilmark{4}, 
J.Borrill\altaffilmark{5,6}, 
A.Boscaleri\altaffilmark{7}, 
K.Coble\altaffilmark{8}, 
B.P.Crill\altaffilmark{9}, 
G.De Gasperis\altaffilmark{1}0, 
P.C.Farese\altaffilmark{8}, 
P.G.Ferreira\altaffilmark{11}, 
K.Ganga\altaffilmark{9,1}2, 
M.Giacometti\altaffilmark{1}, 
E.Hivon\altaffilmark{9}, 
V.V.Hristov\altaffilmark{9}, 
A.Iacoangeli\altaffilmark{1}, 
A.H.Jaffe\altaffilmark{6}, 
A.E.Lange\altaffilmark{9}, 
L.Martinis\altaffilmark{13}, 
S.Masi\altaffilmark{1}, 
P.Mason\altaffilmark{9}, 
P.D.Mauskopf\altaffilmark{14,15}, 
A.Melchiorri\altaffilmark{1}, 
L.Miglio\altaffilmark{16}, 
T.Montroy\altaffilmark{8}, 
C.B.Netterfield\altaffilmark{16}, 
E.Pascale\altaffilmark{7}, 
F.Piacentini\altaffilmark{1}, 
D.Pogosyan\altaffilmark{4}, 
S.Prunet\altaffilmark{4}, 
S.Rao\altaffilmark{17}, 
G.Romeo\altaffilmark{17}, 
J.E.Ruhl\altaffilmark{8}, 
F.Scaramuzzi\altaffilmark{13}, 
D.Sforna\altaffilmark{1}, 
N.Vittorio\altaffilmark{10}
}
\affil{\raggedright
1) Dipartimento di Fisica, Universit\'a di Roma La Sapienza, P.le A. Moro 2, 00185 Roma, Italy,\\
2) Department of Physics, Queen Mary and Westfield College, Mile End Road, London, E1 4NS, UK,\\
3) Jet Propulsion Laboratory, Pasadena, CA, USA,\\
4) CITA University of Toronto, Canada,\\
5) NERSC-LBNL, Berkeley, CA, USA,\\
6) Center for Particle Astrophysics, University of California at Berkeley, 301 Le Conte Hall, Berkeley CA 94720, USA,\\
7) IROE - CNR, Via Panciatichi 64, 50127 Firenze, Italy,\\
8)  Department of Physics, University of California at Santa Barbara, Santa Barbara, CA 93106, USA,\\
9) California Institute of Technology, Mail Code: 59-33, Pasadena, CA 91125, USA,\\
10) Dipartimento di Fisica, Universit\'a di Roma Tor Vergata, Via della Ricerca Scientifica 1, 00133 Roma, Italy,\\
11) Astrophysics, University of Oxford, Keble Road, OX1 3RH, UK,\\
12) PCC, College de France, 11 pl. Marcelin Berthelot, 75231 Paris Cedex 05, France,\\
13) ENEA Centro Ricerche di Frascati, Via E. Fermi 45, 00044 Frascati, Italy,\\
14)  Physics and Astronomy Dept, Cardiff University, UK,\\
15)  Dept of Physics and Astronomy, U.Mass. Amherst, MA, USA,\\
16)  Departments of Physics and Astronomy, University of Toronto, Canada,\\
17)  Istituto Nazionale di Geofisica, Via di Vigna Murata 605, 00143, Roma, Italy,.\\
}

{\bf
The blackbody radiation left over from the Big Bang has been transformed by
the expansion of the Universe into the nearly isotropic 2.73~K Cosmic 
Microwave Background.  Tiny inhomogeneities in the early Universe left 
their imprint on the microwave background in the form of small 
anisotropies in its temperature. These anisotropies contain information 
about basic cosmological parameters, particularly the total energy density 
and curvature of the universe. Here we report the first images of resolved 
structure in the microwave background anisotropies over a significant part 
of the sky.  Maps at four frequencies clearly distinguish the microwave 
background from foreground emission. We compute the angular power spectrum 
of the microwave background, and find a peak at Legendre multipole 
$\ell_{peak}=(197\pm6)$, with an amplitude $DT_{200}=(69\pm8)\mu K$. This is 
consistent with that expected for cold dark matter models in a flat 
(euclidean) Universe, as favoured by standard inflationary scenarios.
}
\\

Photons in the early Universe were tightly coupled to ionized matter through 
Thomson scattering. This coupling ceased about 300,000 years after the 
Big Bang, when the Universe cooled sufficiently to form neutral hydrogen.  
Since then, the primordial photons have travelled freely through the 
universe, redshifting to microwave frequencies as the universe expanded. 
We observe those photons today as the cosmic microwave background (CMB).  
An image of the early Universe remains imprinted in the temperature 
anisotropy of the CMB. Anisotropies on angular scales larger than 
$\sim2^{\circ}$ 
are dominated by the gravitational redshift the photons undergo as they 
leave the density fluctuations present at decoupling$^{(1,2)}$. Anisotropies 
on smaller angular scales are enhanced by oscillations of the photon-baryon 
fluid before decoupling$^{(3)}$. These oscillations are driven by the 
primordial density fluctuations, and their nature depends on the 
matter content of the universe.\\
In a spherical harmonic expansion of the CMB temperature field, 
the angular power spectrum specifies the contributions to the 
fluctuations on the sky coming from different multipoles, each 
corresponding to the angular scale $\theta=\pi/\ell$.  Density fluctuations 
over spatial scales comparable to the acoustic horizon at decoupling 
produce a peak in the angular power spectrum of the CMB, occurring 
at multipole $\ell_{peak}$.  The exact value of $\ell_{peak}$ depends on 
both the linear size of the acoustic horizon and on the angular diameter 
distance from the observer to decoupling.  Both these quantities 
are sensitive to a number of cosmological parameters (see for 
example ref.4), but $\ell_{peak}$ primarily depends on the total density 
of the Universe, $\Omega_0$ . In models with a density $\Omega_0$ near~1, 
$\ell_{peak}\sim 200/\Omega^{1/2}$.  
A precise measurement of peak can efficiently 
constrain the density and thus the curvature of the Universe.
Observations of CMB anisotropies require extremely sensitive and 
stable instruments.  The DMR$^{5}$ instrument on the COBE satellite mapped 
the sky with an angular resolution of $\sim7^{\circ}$, yielding measurements of
the angular power spectrum at multipoles $\ell<20$.  Since then, experiments
with finer angular resolution$^{6-16}$ have detected CMB fluctuations on 
smaller scales and have produced evidence for the presence of a peak 
in the angular power spectrum at peak $\ell_{peak}\sim200$.\\
Here we present high resolution, high signal-to-noise maps of the CMB 
over a significant fraction of the sky, and derive the angular power 
spectrum of the CMB from  $\ell$=~50 to 600. This power spectrum is dominated 
by a peak at multipole $\ell_{peak}=(197\pm6)$ (1$\sigma$ error). 
The existence of this 
peak strongly supports inflationary models for the early universe, 
and is consistent with a flat, Euclidean Universe.
\\

{\bf The Instrumrent}

The BOOMERanG (Balloon Observations Of Millimetric Extragalactic Radiation 
and Geomagnetics) experiment is a microwave telescope that is carried to 
an altitude of $\sim38$~km  by a balloon.  BOOMERanG combines the high 
sensitivity and broad frequency coverage pioneered by an earlier 
generation of balloon-borne experiments with the long  ($\sim10$~days) 
integration time available in a long-duration balloon  flight over 
Antarctica. The data described here were obtained with a focal plane 
array of 16 bolometric detectors cooled to 0.3~K. Single-mode feedhorns 
provide two 18\arcmin~full-width at half-maximum (FWHM) beams at 90~GHz 
and two 10\arcmin~(FWHM) beams at 150~GHz. Four multi-band photometers 
each provide a 10.5', 14' and 13' FWHM beam at 150, 240 and
400~GHz respectively. The average in-flight sensitivity to CMB 
anisotropies was 140, 170, 210 and 2700~$\mu K\cdot s^{1/2}$ at 90, 150, 
240 and 400~GHz respectively. The entire optical system is heavily baffled 
against terrestrial radiation.  Large sun-shields improve rejection 
of radiation from $>60^\circ$ in azimuth from the telescope boresight. 
The rejection has been measured to be greater than 80~dB at all 
angles occupied by the Sun during the CMB observations. Further 
details on the instrument can be found in refs 17-21.
\\

{\bf Observation}

BOOMERanG was launched from McMurdo Station (Antarctica) on 29 December 
1998, at 3:30 GMT. Observations began 3 hours later, and continued 
uninterrupted during the 259-hour flight. The payload approximately 
followed the 79$^\circ$ S parallel at an altitude that varied daily between 
37 and 38.5~km, returning within 50~km of the launch site.\\
We concentrated our observations on a target region, centred at roughly 
right ascension (RA)  5h, declination (dec.) $-45^\circ$, that is uniquely 
free of contamination by thermal emission from interstellar dust$^{(22)}$ 
and that is approximately opposite the Sun during the austral summer. 
We mapped this region by repeatedly scanning the telescope through 60$^\circ$
at fixed elevation and at constant speed. Two scan speeds 
(1$^\circ\cdot$s$ ^{-1}$ and 2$^\circ\cdot$s$ ^{-1}$ in azimuth) 
were used to facilitate tests for systematic effects. 
As the telescope scanned, degree-scale variations in the CMB generated 
sub-audio frequency signals in the output of the detector$^{(23)}$.  
The stability of the detetector system was sufficient to allow 
sensitive measurements on angular scales up to tens of degrees 
on the sky. The scan speed was sufficiently rapid with respect 
to sky rotation that identical structures were observed by detectors 
in the same row in each scan.  Detectors in different rows observed 
the same structures delayed in time by a few minutes.\\
At intervals of several hours, the telescope elevation was interchanged 
between 40$^\circ$, 45$^\circ$ and 50$^\circ$ in order to 
increase the sky coverage and to 
provide further systematic tests. Sky rotation caused the scan centre 
to move and the scan direction to rotate on the celestial sphere. 
A map from a single day at a single elevation covered roughly 22$^\circ$ 
in declination and contained scans rotated by $\pm11^\circ$ on the sky, 
providing 
a cross-linked scan pattern. Over most of the region mapped, each sky 
pixel was observed many times on different days, both at 
1$^\circ\cdot$s$ ^{-1}$ and 2$^\circ\cdot$s$ ^{-1}$
scan speed, with different topography, solar elongation and 
atmospheric conditions, allowing strong tests for any contaminating 
signal not fixed on the celestial sphere.\\
The pointing of the telescope has been reconstructed with an accuracy 
of  2\arcmin~\em r.m.s.\em using data from a Sun sensor and rate gyros.  This 
precision has been confirmed by analyzing the observed positions 
of bright compact HII regions in the Galactic plane (RCW38$^{24}$ , RCW57, 
IRAS08576 and IRAS1022)
and of radio-bright point sources visible in the target region 
(the QSO 0483-436, the BL-Lac object 0521-365 and the blazar 0537-441).
\\

{\bf Calibrations}

The beam pattern for each detector was mapped before flight using a 
thermal source. The main lobe at 90, 150 and 400~GHz is accurately 
modelled by a Gaussian function. The 240~GHz beams are well modelled 
by a combination of two Gaussians. The beams have small shoulders 
(less than 1\% of the total solid angle), due to aberrations in 
the optical system.  The beam-widths were confirmed  in flight 
via observations of compact sources. By fitting radial profiles 
to these sources we determine the effective angular resolution, 
which includes the physical beamwidth and the effects of the 
2\arcmin~\em r.m.s.\em pointing jitter. The effective FWHM angular resoluti
on of the 150~GHz data that we use here to calculate the CMB 
power spectrum is (10$\pm$1)\arcmin, where the error is dominated 
by uncertainty in the pointing jitter.\\
We calibrated the 90, 150 and 240~GHz channels from their measured 
response to the CMB dipole. The dipole anisotropy has been 
accurately (0.7\%) measured by COBE-DMR$^{25}$, fills the beam and 
has the same spectrum as the CMB anisotropies at smaller angular 
scales, making it the ideal calibrator for CMB experiments. 
The dipole signal is typically  $\sim3$~mK peak-to-peak  in each 
60$^\circ$ scan, much larger than the detector noise, and appears in 
the output of the detectors at f=0.008Hz and f=0.016Hz in the 
1$^\circ\cdot$s$ ^{-1}$ and 2$^\circ\cdot$s$ ^{-1}$
scan speeds, respectively. The accuracy of the 
calibration is dominated by two systematic effects: uncertainties 
in the low-frequency transfer function of the electronics, and 
low-frequency, scan-synchronous signals. Each of these is 
significantly different at the two scan speeds. We found that 
the dipole-fitted amplitudes derived from separate analysis 
of the 1$^\circ\cdot$s$ ^{-1}$ and 2$^\circ\cdot$s$ ^{-1}$
data agree to within $\pm$10\% for every 
channel, and thus we assign a 10\% uncertainty in the absolute 
calibration.
\\

{\bf From detector signals to CMB maps}

The time-ordered data comprises $5.4\cdot10^7$ 16-bit samples for 
each channel. These data are flagged for cosmic-ray events, 
elevation changes, focal-plane temperature instabilities, 
and electromagnetic interference events. In general, about 5\% 
of the data for each channel are flagged and not used in the 
subsequent analysis. The gaps resulting from this editing are 
filled with a constrained realization of noise in order to 
minimize their effect in the subsequent filtering of the data.  
The data are deconvolved by the bolometer and electronics 
transfer functions to recover uniform gain at all frequencies.

The noise power spectrum of the data and the maximum-likelihood 
maps$^{26-28}$ are calculated using an iterative technique$^{29}$ that 
separates the sky signal from the noise in the time-ordered data.  
In this process, the statistical weights of frequencies corresponding 
to angular scales larger than 10$^\circ$ on the sky are set to zero to 
filter out the largest-scale modes of the map.  The maps are 
pixelized according to the HEALPix pixelization scheme$^{30}$.

Figure~1 shows the maps obtained in this way at each of the four 
frequencies.  The 400~GHz map is dominated by emission from 
interstellar dust that is well correlated with that observed 
by the IRAS and COBE/DIRBE satellites. The 90, 150 and 240~GHz 
maps are dominated by degree-scale structures that are resolved 
with high signal-to-noise ratio. A qualitative but powerful test 
of the hypothesis that these structures are CMB anisotropy is 
provided by subtracting one map from another. The structures 
evident in all three maps disappear in both the 90$-$150~GHz 
difference and in the 240$-$150~GHz difference, as expected for 
emission that has the same spectrum as the CMB dipole anisotropy 
used to calibrate the maps.

To quantify this conclusion, we performed a "colour index" 
analysis of our data. We selected the $\sim18000$ 14\arcmin~pixel at  
galactic latitude $b<-15^\circ$, and made scatter plots of 90~GHz 
versus 150~GHz and 240~GHz versus 150~GHz. A linear fit to these 
scatter plots gives slopes of $1.00\pm0.15$ and $1.10\pm0.16$, respectively 
(including our present 10\% calibration error), consistent with a 
CMB spectrum. For comparison,  free-free emission with spectral 
index $-$2.35 would produce slopes of 2.3 and 0.85, and is 
therefore rejected with $>99\%$ confidence;  emission from 
interstellar dust with temperature $T_d=15$~K and spectral index 
of emissivity $\alpha=1$ would produce slopes of 0.40 and 2.9. For 
any combination of $T_d >7$~K and $1<\alpha<2$, the dust hypothesis is 
rejected with $>99\%$ confidence. We conclude that the dominant 
source of structure that we detect at 90, 150 and 240~GHz is 
CMB anisotropy. 

We further argue that the 150~GHz map at $b<-15^\circ$ is free of 
significant contamination by any known astrophysical foreground.
Galactic synchrotron and free-free emission is negligible at 
this frequency$^{31}$. Contamination from extra-galactic point sources 
is also small$^{32}$; extrapolation of fluxes from the PMN survey$^{33}$
limits the contribution by point sources (including the three 
above-mentioned radio-bright sources) to the angular power 
spectrum derived below to $<0.7\%$ at $\ell=200$ and $<20\%$ at $\ell=600$. 
The astrophysical foreground that is expected to dominate at 
150~GHz is thermal emission from interstellar dust. We placed 
a quantitative limit on this source of contamination as follows. 
We assumed that dust properties are similar at high ($b<-20^\circ$) 
and moderate ($-20^\circ<b<-5^\circ$) Galactic latitudes. We selected the 
pixels at moderate Galactic latitudes and correlated the structure 
observed in each of our four bands with the IRAS/DIRBE map, 
which is dominated by dust in cirrus clouds. The best-fit slope 
of each of the scatter plots measures the ratios of the dust 
signal in the BOOMERanG channels to the dust signal in the 
IRAS/DIRBE map. We found that the 400~GHz map is very well 
correlated to the IRAS/DIRBE map, and that dust at $b<-20^\circ$)
can account for at most 10\% of the signal variance at 
240 GHz, 3\% at 150~GHz and 0.5\% at 90~GHz.
\\

{\bf Angular Power Spectra}

We compared the angular power spectrum of structures evident 
in Fig.1 with theoretical predictions. In doing so, we separated 
and removed the power due to statistical noise and systematic 
artifacts from the power due to CMB anisotropies in the maps. 
The maximum-likelihood angular power spectrum of the maps was 
computed using the MADCAP$^{34}$ software package, whose algorithms 
fully take into account receiver noise and filtering.

Full analysis of our entire data set is under way.  Because 
of the computational intensity of this process, we report 
here the results of a complete analysis of a limited portion 
of the data chosen as follows.  We analysed the most sensitive 
of the 150~GHz detectors. We restricted the sky coverage to 
an area with $RA>70^\circ$, $b<-20^\circ$ and  
$-55^\circ<Dec<-35^\circ$, and we 
used only the $\sim50\%$ of the data from this detector that 
was obtained at a scan speed of 
1$^\circ\cdot$s$ ^{-1}$.  We used a relatively 
coarse pixelization of 8,000 14-arcmin pixels as a compromise 
between computation speed and coverage of high multipoles. 
Finally, we limited our analysis to $\ell \leq 600$ for which the 
effects of pixel shape and size and our present uncertainty 
in the beam size (1\arcmin) are small and can be accurately modelled. 

The angular power spectrum determined in this way is shown 
in Fig.~2 and reported in Table~1. The power spectrum is 
dominated by a peak at $\ell_{peak}\approx 200$, as predicted by inflationary 
cold dark matter models. These models additionally predict 
the presence of secondary peaks. The data at high $\ell$ limit 
the amplitude, but do not exclude the presence, of a secondary 
peak. The errors in the angular power spectrum are dominated at 
low multipoles ($\ell \leq 350$) by the cosmic/sampling variance, and at 
higher multipoles by detector noise.

The CMB angular power spectrum shown in Fig.2 was derived from 
4.1~days of observation. As a test of the stability of the 
result, we made independent maps from the first and second 
halves of these data. The payload travels several hundred 
kilometers, and the Sun moves 2\arcdeg~on the sky, between these 
maps. Comparing them provides a stringent test for contamination 
from sidelobe pickup and thermal effects. The angular power 
spectrum calculated for the difference map is shown in Fig.2. 
The reduced $\chi^2$ of this power spectrum with respect to zero 
signal is 1.11 (12 degrees of freedom), indicating that the 
difference map is consistent with zero contamination.
\\

{\bf A peak at $\ell \approx 200$ implies a flat Universe}

The location of the first peak in the angular power spectrum of 
the CMB is well measured by by this data setthe BOOMERanG data set. 
From a parabolic fit to the data at $\ell$=50 to 300 in the angular 
power spectrum, we find $\ell_{peak}=(197\pm6)$ (1$\sigma$ error). 
The parabolic fit does not bias the determination of the 
peak molutpolemultipole: applying this method to Monte 
Carlo realizations of theoretical power spectra we recover 
the correct peak location for a variety of cosmological 
models. Finally, Notice the peak location is independent 
of the details of the data calibration, which obviously 
affect only the height of the peak and not its location. 
The height of the peak is 
$\Delta T_{200}=(69\pm4)\pm7$ mK (1-$\sigma$ statistical 
and calibration errors, respectively). 

The data are inconsistent with current models based on 
topological defects (see, for example, ref. 35) but 
are consistent with a subset of cold dark matter models. 
We generated a database of cold dark matter models$^{36,37}$, 
varying six cosmological parameters (the range of 
variation is given in parentheses): the non relativistic 
matter density, $\Omega_m$ (0.05-2);  the cosmological constant, 
$\Omega_{\Lambda}$ (0-1); the Hubble constant, h (0.5-0.8); the baryon 
density, h$^2\Omega_b$ (0.013-0.025), the primordial scalar spectral 
index, ns (0.8-1.3); and the overall normalization A 
(freeparameter) of the primordial density fluctuation 
power spectrum. We compared these models with the power 
spectrum we report here to place constraints on allowed 
regions in this 6-parameter space. In Figure~3 we mark 
with black dots the region of the $\Omega_m-\Omega_\Lambda$ plane where some 
combination of the remaining four parameters within the 
ranges defined by our model space gives a power spectrum 
consistent with our 95\% confidence interval for $\ell_{peak}$. 
This region is quite narrow and elongated along the 
"flat Universe" line $\Omega_m+ \Omega_\Lambda = 1$. The width of this 
region is determined by degeneracy in the models, which 
produce closely similar spectra for different values of 
the parameters$^{38}$. We further evaluated the likelihood of 
the models given the BOOMERanG measurement and the same 
priors (constraints on the values of the cosmological 
parameters) as in ref. 16. Marginalizing over all the 
other parameters, we found the following 95\% confidence 
interval for $\Omega_0 = \Omega_m+\Omega_{\Lambda}$:  
$0.88< \Omega_0 <1.12$. This provides evidence for a euclidean geometry 
of the Universe. Our data clearly show the presence of power 
beyond the peak at $\ell=197$, corresponding to smaller-scale 
structures. The consequences  of this fact will be fully 
analyzed elsewhere.
\\

{\bf Acknowledgments}

The BOOMERanG experiment has been supported by Programma 
Nazionale di Ricerche in Antartide, Universit\'a di Roma
``La Sapienza'', and Agenzia Spaziale Italiana in 
Italy, by NSF and NASA in the USA, and by PPARC in the 
UK. We would like to thank the entire staff of the National 
Scientific Ballooning Facility, and the United States 
Antarctic Program personnel in McMurdo for their excellent 
preflight support and a marvelous LDB flight. Doe/NERSC 
provided the supercomputing facilities.

Correspondence and request for materials should be 
addressed to P.d.B. (e$-$mail: debernardis@roma1.infn.it).
Details on the experiment and numerical data sets are 
available at the web sites 
(http://oberon.roma1.infn.it/boomerang) and 
(http://www.physics.ucsb.edu/$\sim$boomerang).

\begin{table}
{\bf TABLE 1: Angular power specrtum of CMB anisotropy}
\begin{center}
\begin{tabular}{|c|c|c|}
\hline
$\ell-$range & 150~GHz & 150 GHz \\
 & ([$1^{st}$ half]+[$2^{nd}$ half])/2 &([$1^{st}$ half]-[$2^{nd}$ half])/2\\
\hline
[26-75]&1140$\pm$280&63$\pm$32\\ \hline
[76-125]&3110$\pm$490&16$\pm$20\\ \hline
[126-175]&4160$\pm$540&17$\pm$28\\ \hline
[176-225]&4700$\pm$540&59$\pm$44\\ \hline
[226-275]&4300$\pm$460&68$\pm$59\\ \hline
[276-325]&2640$\pm$310&130$\pm$82\\ \hline
[326-375]&1550$\pm$220&-7$\pm$92\\ \hline
[376-425]&1310$\pm$220&-60$\pm$120\\ \hline
[426-475]&1360$\pm$250&0$\pm$160\\ \hline
[476-525]&1440$\pm$290&220$\pm$230\\ \hline
[526-575]&1750$\pm$370&130$\pm$300\\ \hline
[576-625]&1540$\pm$430&-430$\pm$360\\ \hline
\end{tabular}
\end{center}
Shown are measurements of the angular power spectrum 
of the cosmic microwave background at 150~GHz, and test for 
systematic effects. The values listed are for 
$\Delta T_\ell^2=\ell(\ell+1)c_\ell/2\pi$, in mK$^2$.  
Here $c_\ell=<a_{\ell m}^2>$, 
and $<a_{\ell m}^2>$ are the coefficients of the spherical harmonic 
decomposition of the CMB  temperature field:
$\Delta T(\theta,\phi)=\sum a_{\ell m} Y_{\ell m}(\theta,\phi)$.
The stated 1$\sigma$ errors include 
statistical and cosmic/sample variance, and do 
not include a 10\% calibration uncertainty.
\end{table}

\clearpage
\newpage
\begin{figure}
{\bf Figure 1:}
BOOMERanG sky maps (equatorial coordinates). 
The sky maps at 90, 150, and 240~GHz (left panels) are 
shown with a common color scale, using a thermodynamic 
temperature scale chosen such that CMB anisotropies will 
have the same amplitude in the three maps. Only the colour 
scale of the 400~GHz map (bottom right) is 14 times larger 
than the others: this has been done to facilitate comparison 
of emission from interstellar dust (ISD), which dominates 
this map, with ISD emission present in the lower-frequency 
maps. The maps at 90 and 400~GHz are each from a single 
detector, while maps at 150 and 240~GHz have each been 
obtained by co-adding data from three detectors. For purposes 
of presentation, the maps have been smoothed with gaussian 
filters to obtain FWHM effective resolution of 22.5\arcmin (small 
circle in the bottom right side of each panel). Structures 
along the scan direction larger than 10\arcdeg are not present in 
the maps. Several features are immediately evident.  Most 
strikingly, the maps at 90, 150, and 240~GHz are dominated 
by degree-scale structures that fill the map, have well-correlated 
morphology and are identical in amplitude in all three maps. 
These structures are not visible at 400~GHz. The 400~GHz map 
is dominated by diffuse emission which is correlated with the 
ISD emission mapped by IRAS/DIRBE$^{22}$. This emission is 
strongly concentrated towards the right-hand edge of the maps, 
near the plane of the Galaxy.  The same structures are evident 
in the 90, 150 and 240~GHz maps at galactic latitude $b>-15^\circ$, 
albeit with an amplitude that decreases steeply with decreasing 
frequency.\\
The large-scale gradient evident especially near the right 
edge of the 240~GHz map is a result of high-pass filtering 
the very large signals near the Galactic plane (not shown). 
This effect is negligible in the rest of the map. The two 
top right panels show maps constructed by differencing the 
150 and 90~GHz maps and the 240 and 150 GHz~maps. The 
difference maps contain none of the structures that dominate 
the maps at 90, 150 and 240~GHz, indicating that these structures 
do indeed have the ratios of brightness that are unique to the 
CMB. The morphology of the residual structures in the $240-150$~GHz 
map is well-correlated with the 400~GHz map, as is expected if 
the residuals are due to the ISD emission. Three compact sources 
of emission are visible in the lower-frequency maps, as indicated 
by the circles.  These are known radio-bright quasars from the 
SEST pointing catalogue at 230~GHz. The boxed area has been 
used for computing the angular power spectrum shown in Fig.2.
\end{figure}

\begin{figure}
{\bf Figure 2:}
Angular power spectrum measured by BOOMERanG at 150~GHz. 
Each point is the power averaged over $\Delta\ell=50$ and has negligible 
correlations with the adjacent points. The error bars indicate 
the uncertainty due to noise and
cosmic/sampling variance. The errors are dominated by 
cosmic/sampling variance at $\ell<350$; they grow at large $\ell$ 
due to the signal attenuation caused by the combined 
effects$^{39}$ of the 10\arcmin~beam and the 14\arcmin~pixelization 
(0.87 at $\ell$=200 and 0.33 at $\ell$=600). The current $\pm10\%$ uncertainty 
in the calibration corresponds to an overall re-scaling of 
the y-axis by $\pm+20\%$, and is not shown. The current 1\arcmin~uncertainty 
in the angular resolution of the measurement creates an 
additional uncertainty, indicated by the distance between 
the ends of the red error bars and the blue horizontal 
lines, that is completely correlated and is largest (11\%) 
at $\ell = 600$. The green points show the power spectrum of a 
difference map obtained dividing the data in two parts 
corresponding to the first and second halves of the 
timestream. We make two maps (A and B) from these halves, 
and the green points show the power spectrum computed from 
the difference map, (A-B)/2. Signals originating from the 
sky should disappear in this map, so this is a  test for 
contamination in the data (see text). The solid curve has 
parameters 
($\Omega_b,~\Omega_m,~\Omega_\Lambda$, ns, h) 
= (0.05, 0.31, 0.75, 0.95, 0.70) . 
It is the best fit model for the BOOMERanG test flight 
data$^{15,16}$, and is shown for comparison only. The model 
that best fits the new data reported here will be 
presented elsewhere.
\end{figure}

\begin{figure}
{\bf Figure 3:}
Observational constraints on $\Omega_m$ and  $\Omega_\Lambda$.
All the cosmological models (from our data base)
consistent with the position of 
the peak in the angular power spectrum measured by BOOMERanG 
(95\% confidence intervals) define an ``allowed'' region 
in the $\Omega_m-\Omega_\Lambda$ plane (dotted region). Such a region is 
elongated around the $\Omega_0=1$ line identifying a flat 
geometry, Euclidean Universe. The blue lines define 
the age of the Universe for the considered models. 
The green shaded region is consistent (95\% confidence 
contour) with the recent results of the high-redshift 
supernovae surveys$^{40,41}$.
\end{figure}

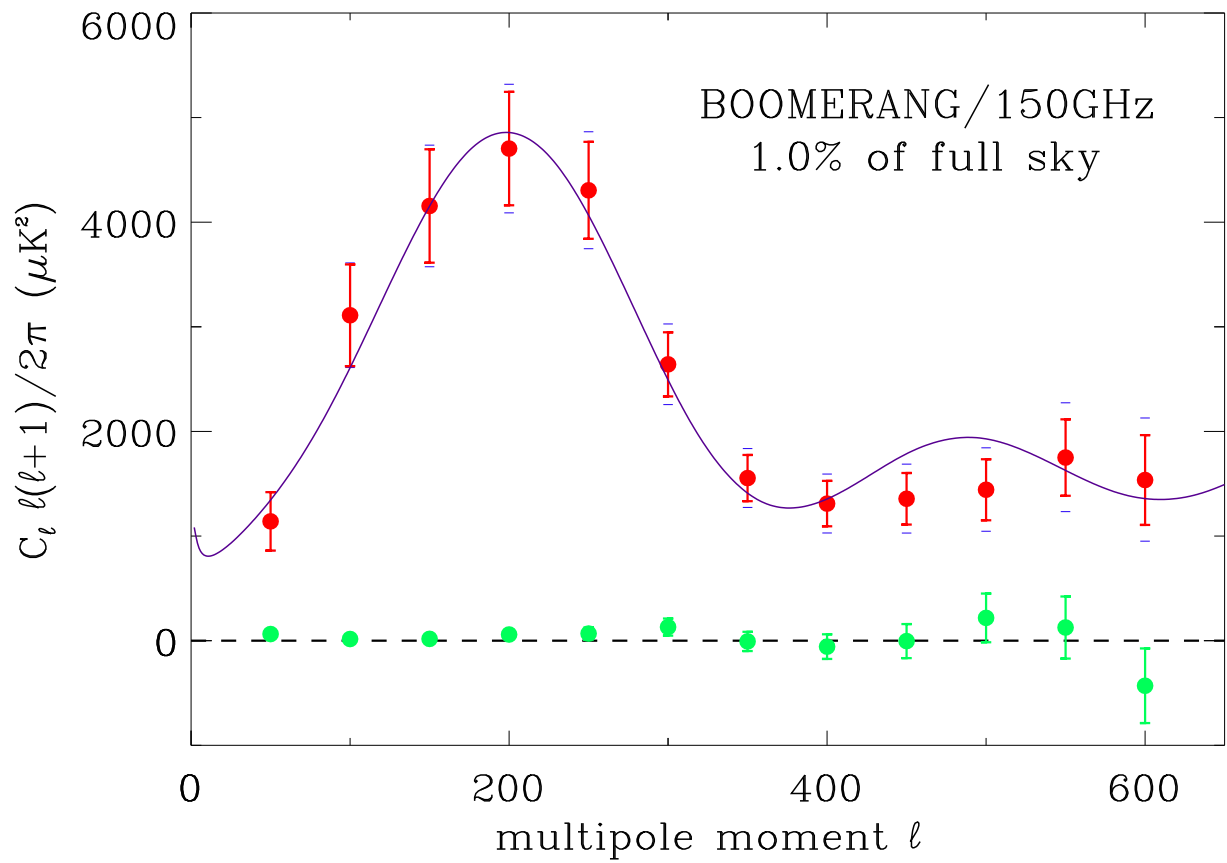
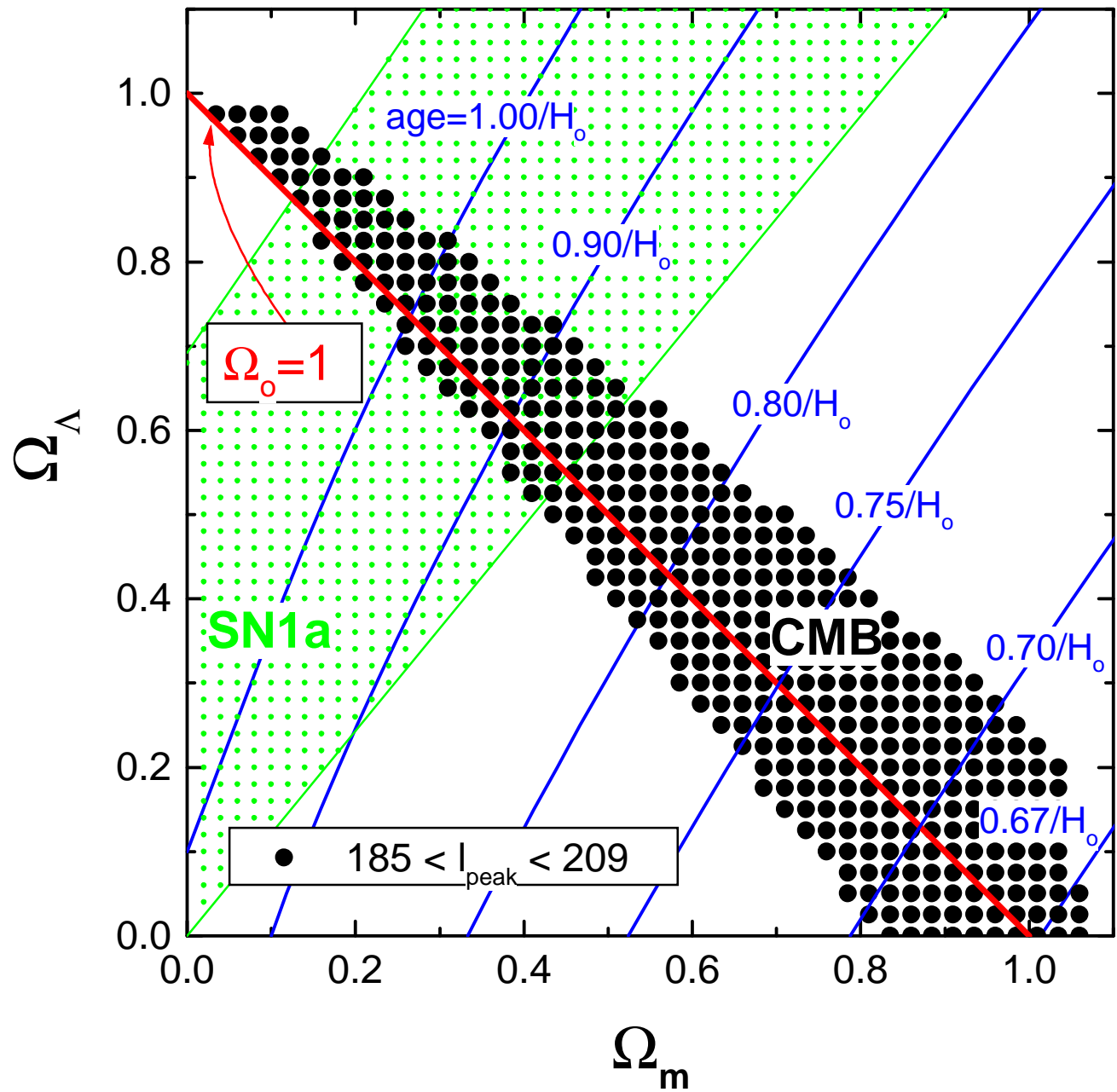
\end{document}